\documentclass[twocolumn,superscriptaddress,prl]{revtex4-1}
\usepackage{amssymb}
\usepackage{amsmath}
\usepackage{graphicx,float,calc}
\usepackage{color,bm}
\usepackage[colorlinks,urlcolor=blue,citecolor=blue,linkcolor=blue]{hyperref}
\usepackage{epstopdf}
\usepackage{amsfonts}
\begin{document}

\title{Observation of non-Hermitian bulk-boundary correspondence in non-chiral non-unitary quantum dynamics of single photons}

\author{Miao Zhang}
\thanks{These authors contribute equally to this work.}
\affiliation{Ministry of Education Key Laboratory for Nonequilibrium Synthesis and Modulation of Condensed Matter,Shaanxi Province Key Laboratory of Quantum Information and Quantum Optoelectronic Devices, School of Physics, Xi'an Jiaotong University, Xi'an 710049, China}

\author{Yue Zhang}
\thanks{These authors contribute equally to this work.}
\affiliation{Ministry of Education Key Laboratory for Nonequilibrium Synthesis and Modulation of Condensed Matter,Shaanxi Province Key Laboratory of Quantum Information and Quantum Optoelectronic Devices, School of Physics, Xi'an Jiaotong University, Xi'an 710049, China}

\author{Shuai Li}
\email{lishuai0999@xjtu.edu.cn}
\affiliation{Ministry of Education Key Laboratory for Nonequilibrium Synthesis and Modulation of Condensed Matter,Shaanxi Province Key Laboratory of Quantum Information and Quantum Optoelectronic Devices, School of Physics, Xi'an Jiaotong University, Xi'an 710049, China}

\author{Rui Tian}
\affiliation{Ministry of Education Key Laboratory for Nonequilibrium Synthesis and Modulation of Condensed Matter,Shaanxi Province Key Laboratory of Quantum Information and Quantum Optoelectronic Devices, School of Physics, Xi'an Jiaotong University, Xi'an 710049, China}

\author{Tianhao Wu}
\affiliation{Ministry of Education Key Laboratory for Nonequilibrium Synthesis and Modulation of Condensed Matter,Shaanxi Province Key Laboratory of Quantum Information and Quantum Optoelectronic Devices, School of Physics, Xi'an Jiaotong University, Xi'an 710049, China}

\author{Yingchao Xu}
\affiliation{Ministry of Education Key Laboratory for Nonequilibrium Synthesis and Modulation of Condensed Matter,Shaanxi Province Key Laboratory of Quantum Information and Quantum Optoelectronic Devices, School of Physics, Xi'an Jiaotong University, Xi'an 710049, China}

\author{Yi-an Li}
\affiliation{Ministry of Education Key Laboratory for Nonequilibrium Synthesis and Modulation of Condensed Matter,Shaanxi Province Key Laboratory of Quantum Information and Quantum Optoelectronic Devices, School of Physics, Xi'an Jiaotong University, Xi'an 710049, China}

\author{Yuanbang Wei}
\affiliation{Ministry of Education Key Laboratory for Nonequilibrium Synthesis and Modulation of Condensed Matter,Shaanxi Province Key Laboratory of Quantum Information and Quantum Optoelectronic Devices, School of Physics, Xi'an Jiaotong University, Xi'an 710049, China}

\author{Hong Gao}
\affiliation{Ministry of Education Key Laboratory for Nonequilibrium Synthesis and Modulation of Condensed Matter,Shaanxi Province Key Laboratory of Quantum Information and Quantum Optoelectronic Devices, School of Physics, Xi'an Jiaotong University, Xi'an 710049, China}

\author{M. Suhail Zubairy}
\affiliation{Institute for Quantum Science and Engineering (IQSE) and Department of Physics and Astronomy, Texas A\&M University, College Station, TX 77843-4242, USA}

\author{Fuli Li}
\affiliation{Ministry of Education Key Laboratory for Nonequilibrium Synthesis and Modulation of Condensed Matter,Shaanxi Province Key Laboratory of Quantum Information and Quantum Optoelectronic Devices, School of Physics, Xi'an Jiaotong University, Xi'an 710049, China}

\author{Bo Liu}
\email{liubophy@gmail.com}
\affiliation{Ministry of Education Key Laboratory for Nonequilibrium Synthesis and Modulation of Condensed Matter,Shaanxi Province Key Laboratory of Quantum Information and Quantum Optoelectronic Devices, School of Physics, Xi'an Jiaotong University, Xi'an 710049, China}

\begin{abstract}
{\bf The breakdown of conventional bulk-boundary correspondence, a cornerstone of topological physics, is one of counter-intuitive phenomena in non-Hermitian systems, that is deeply rooted in symmetry. In particular, preserved chiral symmetry is one of the key ingredients, which plays a pivotal role in determining non-Hermitian topology. Nevertheless, chiral symmetry breaking in non-Hermitian systems disrupts topological protection, modifies topological invariants, and substantially reshapes spectral and edge-state behavior. The corresponding fundamentally important bulk-boundary correspondence thus needs to be drastically reconstructed. However, it has so far eluded experimental efforts. Here, we theoretically predict and experimentally demonstrate the bulk-boundary correspondence of a one-dimensional (1D) non-Hermitian system with chiral symmetry breaking in discrete-time non-chiral non-unitary quantum walks of single photons. Through constructing a domain-wall configuration, we experimentally observe the photon localization at the interface of domain-wall structure, clearly indicating the presence of the topological edge mode. The appearance of that matches excellently with the prediction of our introduced non-chiral non-Bloch topological invariants pair. Our work thus unequivocally builds the non-Hermitian bulk-boundary correspondence as a general principle for studying topological physics in non-Hermitian systems with chiral symmetry breaking.}
\end{abstract}

\maketitle
The bulk-boundary correspondence (BBC) is a fundamental principle in exploring topological states of matter~\cite{Moessner2021, Hasan2010, Qi2011}, where the existence of anomalous boundary modes under the open boundary condition (OBC) can be faithfully captured by the bulk topology. Such a conventional BBC implicitly assumes that the system under OBC shares essentially the same bulk eigenvalue spectra with its counterpart under the periodic boundary condition (PBC). However, this assumption no longer holds in a broad class of non-Hermitian systems and the abnormal breakdown of conventional BBC is thus induced by the non-Hermiticity~\cite{Yao2018, Kunst2018, Lee2016, Leykam2017, Xiong2018, Shen2018, Martinez Alvarez2018, Yao22018, Lee2019, Jin2019, Herviou2019, Zirnstein2021, Pocock2019, Borgnia2020, Yokomizo2019, Brzezicki2019, Yang2020, Kawabata2020, Zirnstein22021, Helbig2020, Weidemann2020}. To comprehensively understand it, the non-Bloch band theory has been established~\cite{Yao2018, Kunst2018, Yao22018, Yokomizo2019, Yang2020}. Based on a generalized Brillouin zone (GBZ), non-Bloch topological invariants are introduced and BBC can be restored. Recent experimental progresses in manipulating the non-Hermiticity in synthetic systems provide unprecedented opportunities for investigating the non-Hermitian BBC~\cite{Weidemann2020, Brandenbourger2019, Helbig2020, Xiao2020, Zhou2023, Zhan2017, Liang2022}. It is shown that
symmetries, in particular, chiral symmetry, being one fundamental symmetry,
play a pivotal role in determining non-Hermitian BBC~\cite{Kawabata2019,Gong2018}. As distinct from non-Hermitian systems preserving the chiral symmetry~\cite{Xiao2020, Zhou2023, Zhan2017}, breaking such symmetry disrupts topological protection, modifies topological invariants, and substantially reshapes eigenvalue spectra and edge states. The corresponding fundamentally important BBC thus needs to be drastically
reconstructed. However, it has never been experimentally demonstrated yet in any system.

In this work, we theoretically predict and experimentally demonstrate the BBC of a 1D non-Hermitian system with chiral symmetry breaking, implemented in discrete-time non-chiral non-unitary quantum walks of single photons. To reconstruct the corresponding  non-Hermitian BBC, we first introduce the non-chiral non-Bloch topological invariants pair. Interestingly, we find that such topological invariants pair not only can faithfully
identify the emergent topological exceptional (EP) region, a new intrinsic non-Hermitian feature being absent in the chiral case~\cite{Xiao2020, Zhou2023, Zhan2017}, but it also can correctly predict the existence of topological edge modes in both emergent EP and gapped regions uniformly. To experimentally demonstrate that, a domain-wall configuration is constructed. We observe that photons become dynamically localized at the interface of domain-wall structure when topological edge states exist. Amazingly, such experimentally measured topological edge states match excellently with the prediction of our introduced non-chiral non-Bloch invariants pair. Therefore, the fundamentally important BBC in the non-Hermitian system with chiral symmetry breaking is established, that should  pave the way for future studies of topological effects in non-chiral non-Hermitian systems.

{\bf Results}

{\bf Non-chiral non-unitary quantum walk.} A 1D non-chiral non-unitary quantum walk studied here can be captured by the following Floquet operator
\begin{equation}
U_{0}=T_{\downarrow}R_{y}(\theta_{2})MT_{\uparrow}R_{y}(\theta_{1}),
\label{eq:U0}
\end{equation}
where $T_{\uparrow}(T_{\downarrow})$ stands for the pseudospin-dependent translation by one lattice site, where $\uparrow$ and $\downarrow$ refer to the coin state $\vert 0\rangle$ and $\vert 1\rangle$, respectively. $R_{y}(\theta)$ labels the coin operator, referring to rotate coin states by $\theta$ around y-axis. $M=\mathbb{I}_{x}\otimes \left(e^{\gamma}\vert 0 \rangle \langle 0 \vert + e^{-\gamma}\vert 1 \rangle \langle 1\vert \right)$ is the polarization selective loss operator, introducing the non-unitarity with $\gamma$ being a tunable parameter in experiments. The Floquet system captured by $U_{0}$ corresponds to an effective non-Hermitian Hamiltonian ${\mathbf H}_{\rm eff}$ through the relation $U_{0}=e^{-i{\mathbf H}_{\rm eff}}$. Following the non-Bloch band theory~\cite{Yao2018, Kunst2018, Yao22018, Yokomizo2019, Yang2020}, ${\mathbf H}_{\rm eff}$  can be expressed on GBZ. By replacing the Bloch phase factor $e^{i k}$ with $\beta$ on GBZ, ${\mathbf H}_{\rm eff}$ can be expressed as $\tilde{\bf{h}}(\beta) \cdot {\bf\sigma}$ with ${\bf\sigma}$ being the Pauli vector and
\begin{equation}
\begin{aligned}
&\tilde{\bf{h}}_{x}=-\frac{E}{2i\sin E}(\beta \alpha -\beta^{-1}\alpha^{-1})\cos \theta_{2}/2 ,\\
&\tilde{\bf{h}}_{y}=\frac{E}{2i\sin E}[\ i(\alpha^{-1} +\alpha)\cos \theta_{1}/2 \sin \theta_{2}/2 \\
& \ \ \ + i(\beta \alpha +\beta^{-1}\alpha^{-1})\sin \theta_{1}/2 \cos \theta_{2}/2 \ ], \\
&\tilde{\bf{h}}_{z}=\frac{E}{2i\sin E}(\alpha -\alpha^{-1})\sin \theta_{2}/2, \\
\end{aligned}
\label{eq:h}
\end{equation}
where $E \equiv E_{\pm} =\pm \arccos (-\frac{\alpha+\alpha^{-1}}{2} \sin \theta_{1}/2 \sin \theta_{2}/2 + \frac{\beta \alpha + \beta^{-1} \alpha^{-1}}{2} \cos \theta_{1}/2 \cos \theta_{2}/2)$ and $\alpha=e^{-\gamma}$. As distinct from the Hermitian case with $\alpha=1$, where the relation $\Gamma {\mathbf H}_{\rm eff} \Gamma^{-1}=-{\mathbf H}_{\rm eff}$ with $\Gamma=\sigma_{z}$ is satisfied, such a relation can not be held when considering $\alpha\neq1$. Therefore, the chiral symmetry is broken here due to the presence of the non-Hermiticity ($\alpha\neq 1$) and a non-chiral non-unitary quantum walk is thus achieved.

{\bf The non-chiral non-Bloch invariants pair.} To correctly capture topological properties of $U_0$, we introduce the concept of non-chiral non-Bloch invariants pair. Through rewriting $U_{0}$ in two different time frames as
\begin{equation}
\begin{aligned}
 &U'=R_{y}(\theta_{1}/2)T_{\downarrow}R_{y}(\theta_{2})MT_{\uparrow}R_{y}(\theta_{1}/2), \\ &U''=R_{y}(\theta_{2}/2)MT_{\uparrow}R_{y}(\theta_{1})T_{\downarrow}R_{y}(\theta_{2}/2),
 \end{aligned}
\end{equation}
the non-chiral non-Bloch invariants pair can be defined by adding the generalized boundary condition (GBC)~\cite{Zhang2024, Guo2021} to $U'$ and $U''$. Topological invariants associated with $U'$ and $U''$ can thus be obtained, for instance,
\begin{equation}
\nu'=-\frac{1}{\pi}\sum\nolimits_{\pm}\oint_{C^{'{\rm inside}}_{\beta}(\bigtriangleup \rightarrow 0)}d \beta \langle \psi_{\pm}^{'L} \vert i \partial_{\beta} \vert \psi_{\pm}^{'R} \rangle,
\label{eq:nu'}
\end{equation}
with $\vert \psi_{\pm}^{'R} \rangle$ and $\langle \psi_{\pm}^{'L} \vert$ being right and left eigenstates of $U'$, respectively. $C^{'{\rm inside}}_{\beta}$ stands for the corresponding GBZ (inner loop) of $U'$ under GBC (see details in Supplemental Information(SI)). Back to the original time frame, we can introduce  topological invariants pair from $\nu'$ and $\nu''$ to characterize the bulk topology of $U_0$, which can be expressed as
\begin{equation}
\nu_{0}=\frac {\nu'+\nu''}{2}, \ \nu_{\pi}=\frac {\nu'-\nu''}{2}.
\label{eq:nu1}
\end{equation}

\begin{figure}[tbp]
\includegraphics[width=0.45\textwidth]{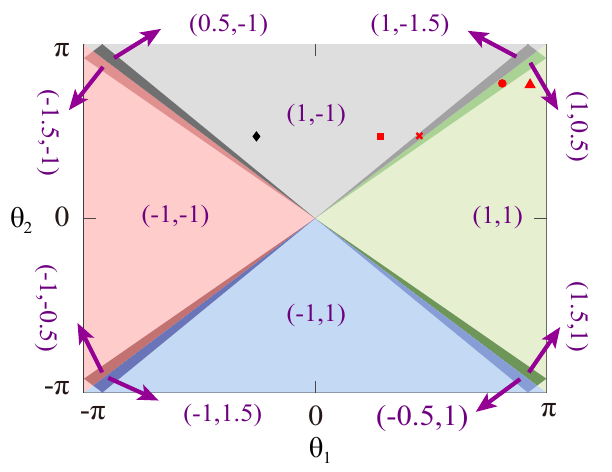}
\caption{{\bf Topological phase diagram of the 1D non-chiral non-unitary quantum walk governed by $U_0$.} Different topological phases  characterized by distinct topological invariants pair $(\nu_{0}, \nu_{\pi})$  as the function of coin parameters
$({\theta _1}, {\theta _2})$ when fixing $\gamma=-\frac{1}{4}\ln \eta$ with $\eta=0.64$. The gapped regions are characterized with integer topological invariants pair, while exceptional regions are captured with one of topological invariants pair being a half integer. Different colored symbols indicate coin parameters for constructing various domain wall structures to observe topological edge states.}
\label{fig:fig1}
\end{figure}

\begin{figure*}[tbp]
\includegraphics[width=0.8\textwidth]{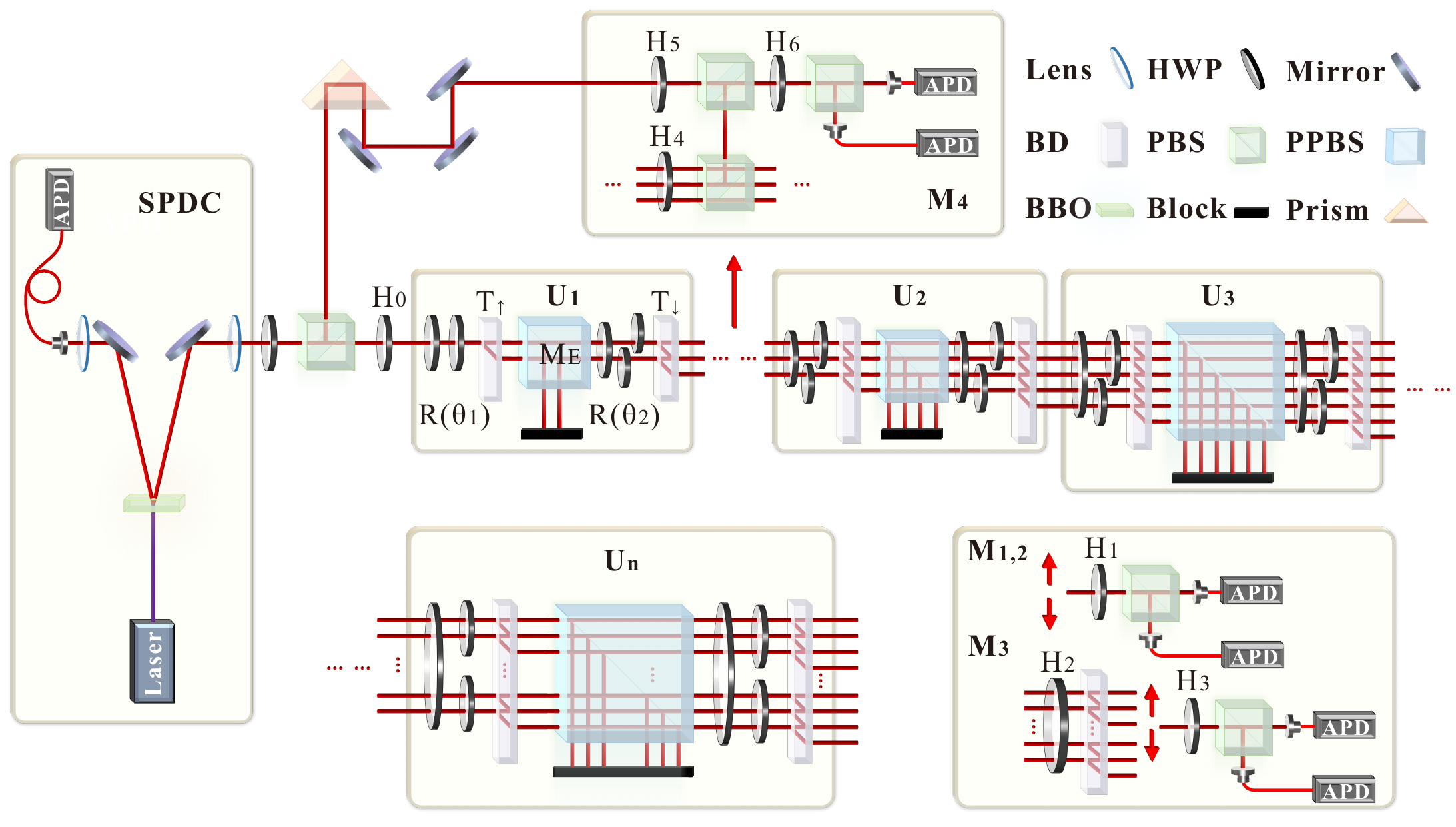}
\caption{{\bf Experimental setup.} Heralded single photons are produced by type-I spontaneous parametric down conversion. One photon servers as the trigger and the other is used as the
signal. Signal photons then proceed through the quantum-walk interferometric network, where coin operators, shift operators and selective-loss operator are  implemented by a set of half wave plates (HWPs), beam displacers (BDs) and  partially polarizing beam splitters (PPBSs), respectively. To construct the time-integrated wave function of photons,
four types of projective measurements with (M$_4$) and without (M$_{1,2,3}$) interfering with the initial photon state are conducted. Finally, the photon is detected by avalanche photodiodes (APDs) in coincidence with the trigger.}
\label{fig:fig2}
\end{figure*}

\begin{figure}[tbp]
\includegraphics[width=0.45\textwidth]{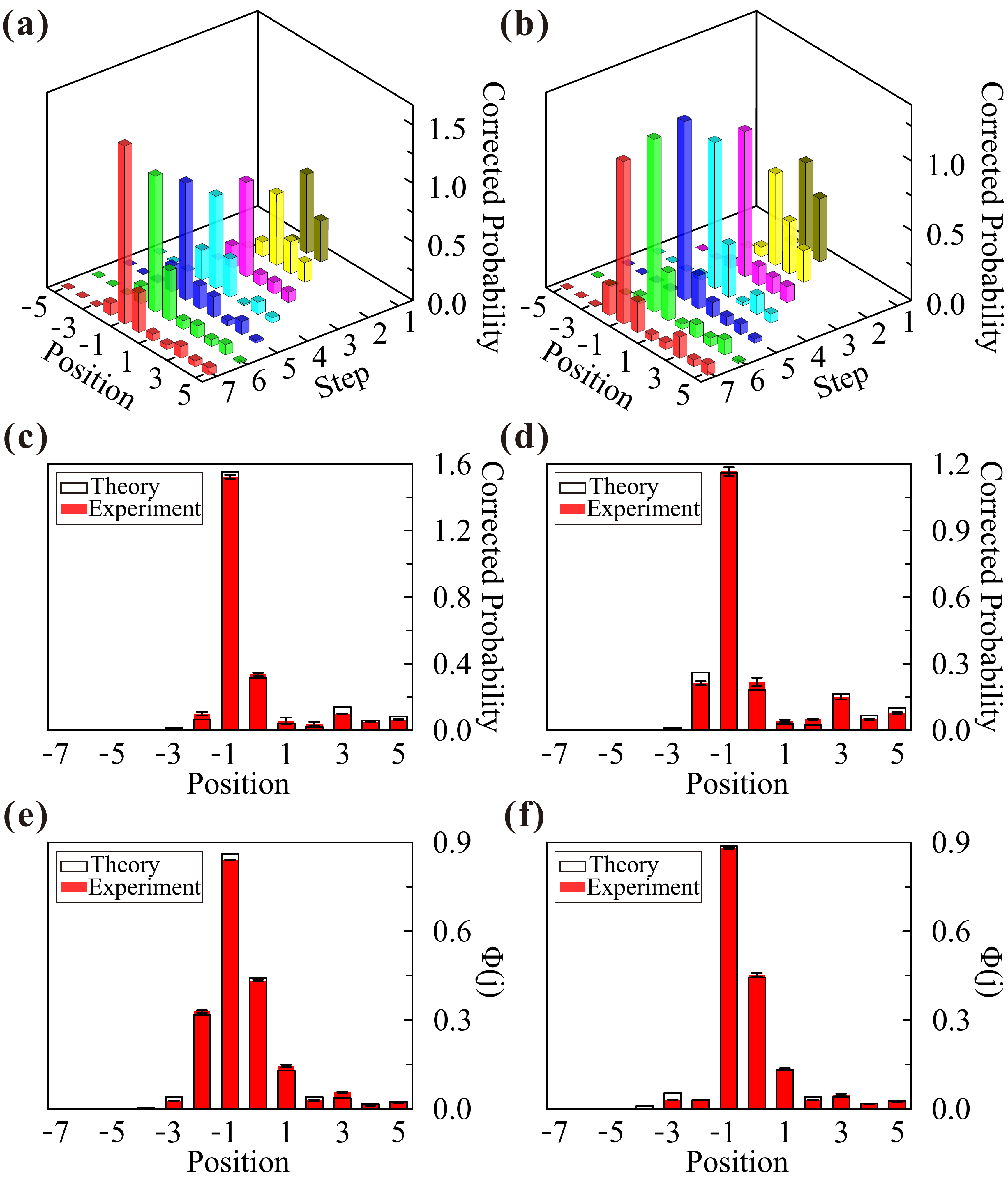}
\caption{{\bf Observation of the topological edge mode at the interface of domain structure.} {(a)} and {(b)} Measured spatial probability distributions of a seven-step non-chiral non-unitary quantum walk for different domain structures. {(c)} and {(d)} are the corresponding distribution at the last step. {(e)} and {(f)} Spatial distribution $\Phi(j)$ of the time-integrated wave function after a seven-step quantum walk. Coin parameters for the left and right columns are different, where the left segments are chosen with  ($\theta_{1}^{L}=0.95\pi$, $\theta_{2}^{L}=0.8\pi$) and ($\theta_{1}^{L}=0.85\pi$, $\theta_{2}^{L}=0.8\pi$), respectively, and the right segments are the same with
($\theta_{1}^{R}=-0.3\pi$, $\theta_{2}^{R}=0.5\pi$). The  initial state is chosen as $\vert j=-1 \rangle \otimes \frac{1}{\sqrt{2}}(\vert 0 \rangle - \vert 1 \rangle)$. Other parameters are the same as in Fig. 1.}
\label{fig:fig5}
\end{figure}
In Fig.~\ref{fig:fig1}, we show the  non-Hermitian topological phase diagram on $\theta_{1}-\theta_{2}$ plane characterized with $(\nu_{0}, \nu_{\pi})$. As distinct from the chiral case~\cite{Xiao2020, Zhou2023, Zhan2017}, topologically nontrivial exceptional regions (regions possessing EP points) emerge, which is a new intrinsic non-Hermitian feature, being absent in the chiral case. Interestingly, we find that half-integer characters of our introduced topological invariants, missing in the chiral case~\cite{Xiao2020, Zhou2023, Zhan2017}, can faithfully identify the emergent exceptional region.  Specifically, we find that half integer difference $\bigtriangleup \nu_{0} \ (\bigtriangleup \nu_{\pi})$ between two regions indicates the emergence of gap closing at  quasienergy $0 \ (\pi)$ and thus captures the phase boundary (see details in SI). Furthermore, $\nu_{0} \ (\nu_{\pi})$ can also correctly predict the existence of zero- or $\pi$-edge topological mode in distinct regions and thus are capable to reconstruct the corresponding BBC. Although quasienergy is not real here, the quasienergy Brillouin zone can still be well defined in its real part, deemed as a periodic phase. Therefore, the zero($\pi$)-edge mode~\cite{Rudner2013, Zhou2014} can be defined by the real part of quasienergy. When considering a domain-wall structure composited with two segments in gapped regions characterized with distinct integer $\nu_{0} \ (\nu_{\pi})$, the existence or absence of topological $0$ ($\pi$)-edge mode can be correctly predicted by whether $\bigtriangleup \nu_{0} \ (\bigtriangleup \nu_{\pi})$  being nonvanishing or not. More interestingly, we also find that $\nu_{0} \ (\nu_{\pi})$ can also amazingly predict the correct BBC in exceptional regions. To understand that, taking $\nu'$ as an example, we can rewrite it as
\begin{equation}
\nu'=\frac {1}{\pi} \oint_{C'^{\rm inside}_{\beta}(\bigtriangleup \rightarrow 0)}(d \theta' + d \phi')\equiv \tilde{\nu}'+\tilde{\tilde{\nu}}' ,
\label{eq:nu'2}
\end{equation}
where $e^{i\theta^{\prime}}=\frac{h_{x}^{\prime}+ih_{y}^{\prime}}{\left\vert
h_{x}^{\prime}+ih_{y}^{\prime}\right\vert }$ and $\tan2\phi^{\prime}=\frac{(\sin
E-\sin E^{\ast})h_{z}^{\prime}}{\sin E+\sin E^{\ast}}$ with
$h_{x}^{\prime}=-\frac{E}{2i\sin E}(\alpha\beta_{n}-\alpha^{-1}\beta_{n}^{-1}%
)\cos\theta_{2}/2$, $h_{y}^{\prime} =\frac{E}{2i\sin E}[i(\alpha\beta_{n}+\alpha^{-1}\beta_{n}^{-1}%
)\cos\theta_{2}/2\sin\theta_{1}/2+i(\alpha^{-1}+\alpha)\sin\theta_{2}/2\cos\theta
_{1}/2]$ and $h_{z}^{\prime}    =\frac{E}{2i\sin E}(\alpha-\alpha^{-1})\sin\theta_{2}/2$. $\tilde{\nu}'$ in Eq.~\ref{eq:nu'2} can be viewed as a winding number which accounts for times of trajectories of eigenvectors of $U'$ on GBZ passing around the $z$-axis (see SI for details). It thus defines the geometrical meaning of $\nu'$ and manifests its topological nature. $\tilde{\tilde{\nu}}'$ in Eq.~\ref{eq:nu'2} captures the corresponding gap closing between gapped and exceptional regions. In the gapped region, $\tilde{\tilde{\nu}}'$ is zero. While in the exceptional region, it becomes a half integer. Therefore, when considering a domain-wall structure, the existence of topological $0$ ($\pi$)-edge mode can be correctly predicted by $\bigtriangleup \tilde{\nu}_{0} \ (\bigtriangleup \tilde{\nu}_{\pi})$ between two segments.

{\bf Experimental demonstration of the  Non-Hermitian BBC.} We experimentally investigate the non-chiral non-unitary quantum-walk dynamics governed by $U_{0}$ through the implementation of photonic discrete-time quantum walk (DTQW) via the spatial-mode multiplexing scheme~\cite{Broome2010}. The experimental setup is illustrated in Fig.~\ref{fig:fig2}. Here, coin states are realized by horizontally and vertically polarized photons, respectively, labeled by $\vert 0 \rangle$ and $\vert 1 \rangle$, correspondingly. Spatial modes of photons represent lattice sites. The polarization-dependent translation operator is implemented through using  beam displacers (BDs), where the horizontally polarized photon is displaced to the neighboring spatial mode, while the vertically polarized photon remains unchanged. The $M$ operator is implemented by using the partially polarizing beam splitter (PPBS). In experiments, we set the loss parameter $\gamma=-\frac{1}{4}\ln \eta$ with $\eta$ fixed at $0.64$, achieved by using the PPBS with a selected polarization-dependent transmissivity (See Methods for details).

To investigate the non-Hermitian BBC in our non-unitary quantum walks without chiral symmetry, we consider a domain-wall structure, where the interface is located between the lattice site $j=0$ and $j=-1$, separating the right bulk ($0 \leq j \leq N$) and left bulk ($-N \leq j \leq -1$). Distinct coin parameters ($\theta^{L,R}_{1}, \theta^{L,R}_{2}$) are assigned to the corresponding segments, respectively. Here, we construct the domain-wall structure by fixing the right bulk with ($\theta_{1}^{R}=-0.3\pi$, $\theta_{2}^{R}=0.5\pi$) and varying the left bulk through selecting different regions in the phase diagram (diamond standing for the right segment and other patterns for different left segments as shown in Fig.~\ref{fig:fig1}). A seven-step quantum-walk is conducted under the above domain-wall configurations. Initializing the photon at $j=-1$, but with different coin states, the spatial photon distribution after corrected via
multiplying by a factor $e^{2\gamma t}$
due to the difference between $M_E$ and $M$ (See Methods for details)~\cite{Xiao2020} has been obtained, for instance, as shown in Fig. 3--Fig. 5. Specifically, \textit{first}, we set the left bulk by selecting two distinct gapped regions with {($\theta_{1}^{L}=0.3\pi$, $\theta_{2}^{L}=0.5\pi$) and ($\theta_{1}^{L}=0.95\pi$, $\theta_{2}^{L}=0.8\pi$}), respectively. For both cases, $\bigtriangleup \nu_{0}$ between two segments are zero, but as regards $\bigtriangleup \nu_{\pi}$, for the former case it is zero, but for the later case, it is nonvanishing. $\bigtriangleup \nu_{\pi} \neq 0$ ($\bigtriangleup \nu_{\pi} = 0$) indicates the appearance (absence) of topological $\pi$-edge mode existed at the interface between $j=0$ and $j=-1$. From numerics, we find that for the considered domain structure, the topological $\pi$-mode at the interface has the largest imaginary part of the eigenenergy. Therefore, such a topological edge mode can be identified through the dynamical evolution. In experiments, we initialize the photon at the interface and let it evolve for
seven-step quantum walk. To extract the $\pi$-edge mode here, we utilize the so-called weighted summation method~\cite{Xiao2020} to construct the time-integrated wave function as follows
\begin{equation}
\vert \Phi(t) \rangle=\sum_{t'=0}^{t} \frac{e^{i \pi t'}} {t+1} \vert \phi(t')\rangle,
\label{eq:varPhipi}
\end{equation}
with $\vert \phi(t)\rangle$ being the time-evolved wave function of photons. The above weighted summation at long time can clearly select out the $\pi$-edge mode at the interface because it has the largest imaginary part of the eigenenergy (see SI for details). Through  measuring $\Phi(j)=\sqrt{\vert(\langle j \vert \otimes \langle 0 \vert)\vert \Phi(t) \rangle \vert^{2}+\vert(\langle j \vert \otimes \langle 1\vert)\vert \Phi(t) \rangle \vert^{2}}$. As shown in {Fig.~\ref{fig:fig3}(c)}, a prominent peak exists at the interface for the domain structure with $\bigtriangleup \nu_{\pi} \neq 0$, clearly indicating the presence of the topological $\pi$-edge mode. While for the case with $\bigtriangleup \nu_{\pi} = 0$ , as shown in Fig.~\ref{fig:fig3}(d), the wavefunction spreads out during the dynamical evolution, manifesting the absence of topological edge mode at the interface.

\begin{figure}[tbp]
\includegraphics[width=0.45\textwidth]{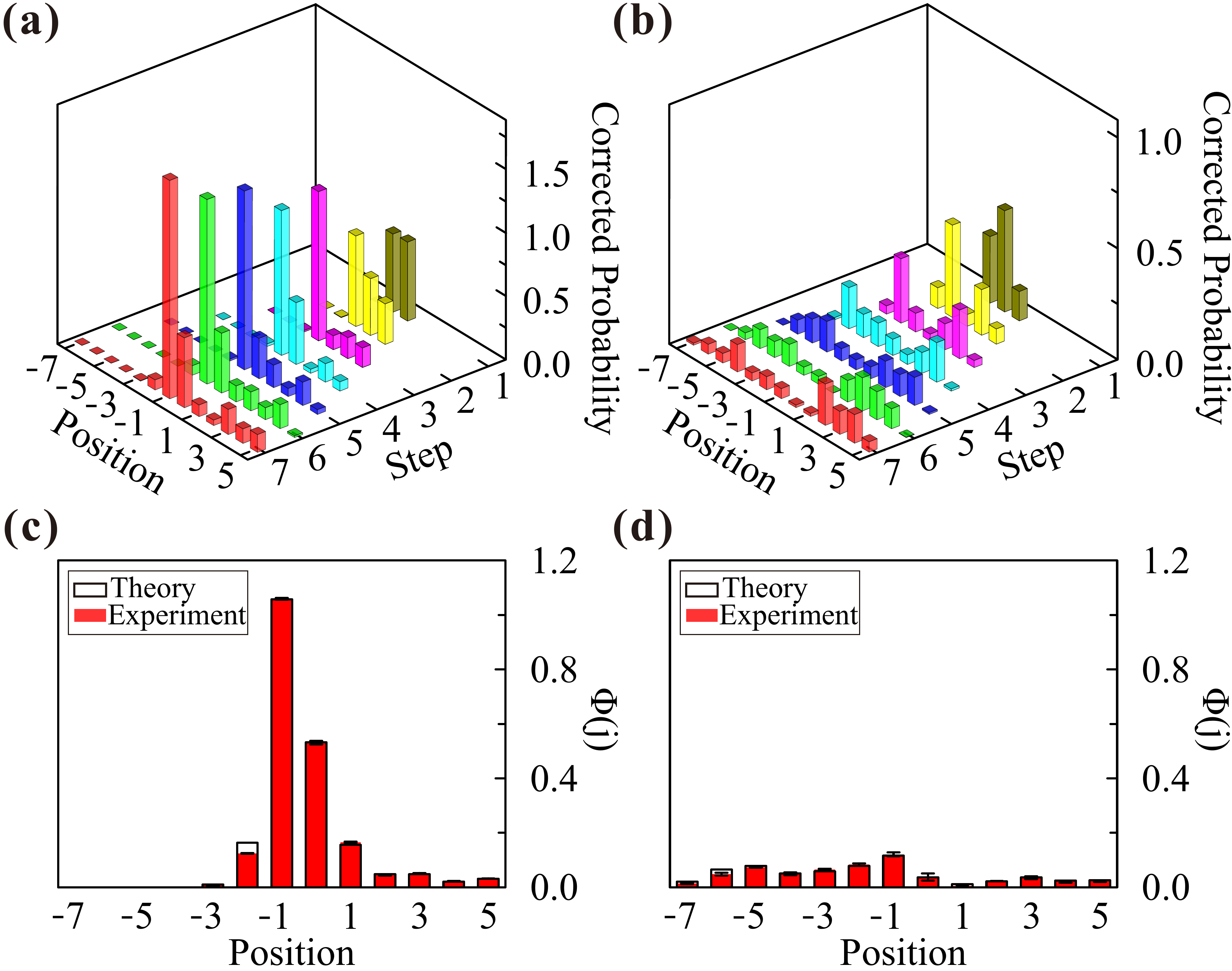}
\caption{{\bf Observation of the non-Hermitian BBC in gapped regions.} {(a)} and {(c)} Measured spatial probability distribution and spatial distribution of  the time-integrated wave function in
a seven-step quantum walk. Other parameters are chosen as the same in the left column of Fig. 3. For comparison, { (b)} and {(d)} show the case with different left segment  ($\theta_{1}^{L}=0.3\pi$, $\theta_{2}^{L}=0.5\pi$). The initial state is chosen as $\vert j=-1 \rangle \otimes \vert 1 \rangle$.}
\label{fig:fig3}
\end{figure}

\begin{figure}[tbp]
\includegraphics[width=0.45\textwidth]{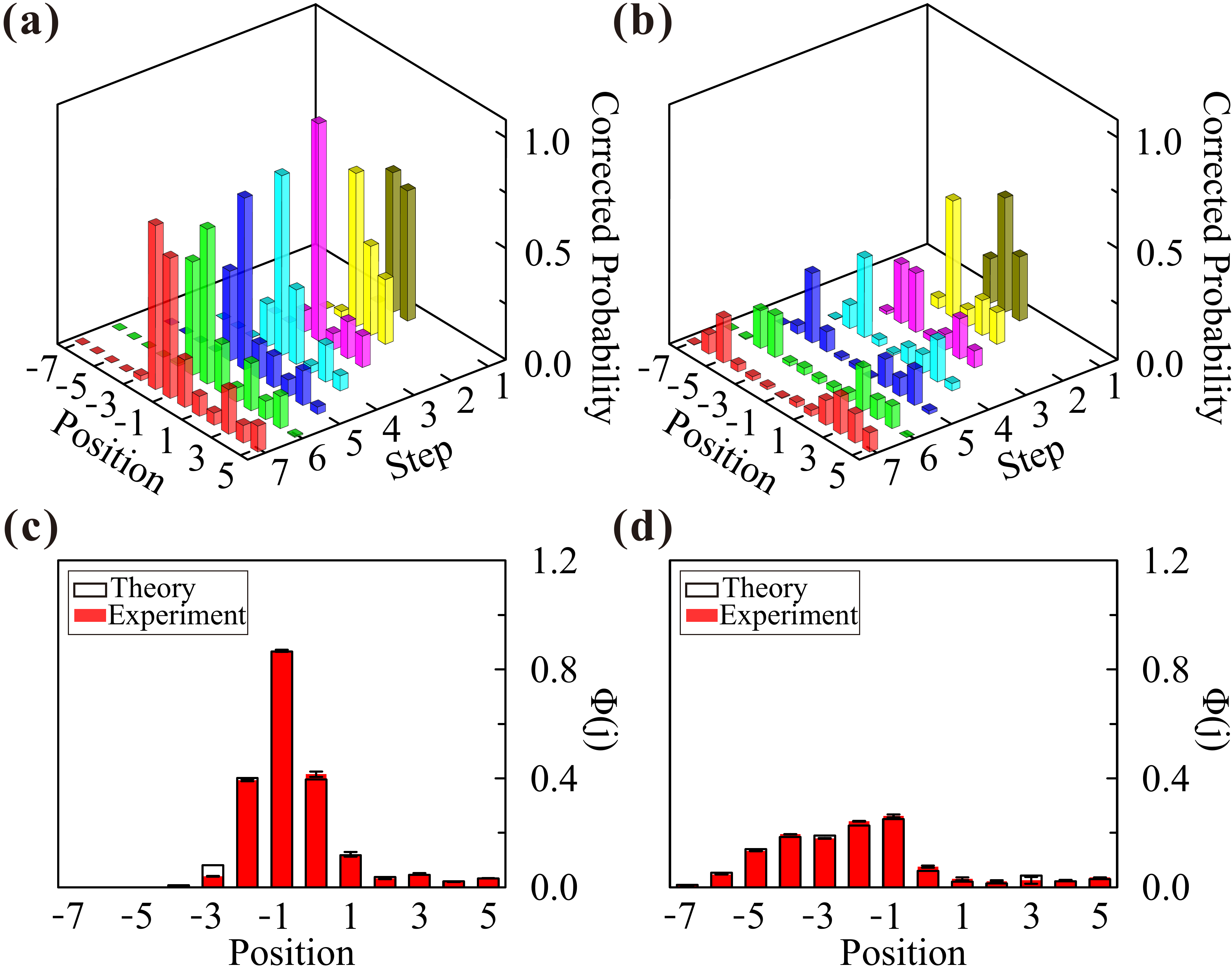}
\caption{{\bf  Observation of the non-Hermitian BBC in exceptional regions.} {(a)} and {(c)} Measured spatial probability distribution and spatial distribution of the time-integrated wave function in
a seven-step quantum walk. Other parameters are chosen as the same in the right column of Fig. 3. For comparison, {(b)} and {(d)} show the case with  different left
segment ($\theta_{1}^{L}=0.47\pi$, $\theta_{2}^{L}=0.5\pi$). The initial state is the same as in Fig. 4.}
\label{fig:fig4}
\end{figure}
\textit{Second}, we set the left segment by selecting two distinct exceptional regions with
($\theta_{1}^{L}=0.85\pi$, $\theta_{2}^{L}=0.8\pi$) and ($\theta_{1}^{L}=0.47\pi$, $\theta_{2}^{L}=0.5\pi$), respectively. For both cases,  $\bigtriangleup \tilde{\nu}_{0}$ between two segments are zero, but as regards $\bigtriangleup \tilde{\nu}_{\pi}$, for the former case it is non-zero, but for the later it is zero.
$\bigtriangleup \tilde{\nu}_{\pi}\neq 0$ indicates that the topological $\pi$-edge mode exist at the interface between $j=0$ and $j=-1$. As shown in Fig.~\ref{fig:fig4}(c), a prominent peak exists at the interface. While for $\bigtriangleup \tilde{\nu}_{\pi}=0$, such a peak disappears, as shown in Fig.~\ref{fig:fig4}(d), indicating the absence of topological $\pi$-edge mode. Therefore, the measured topological edge modes at the interface of domain structures are correctly predicted by our defined non-chiral non-Bloch invariants pair, which unambiguously reveals the corresponding non-Hermitian BBC.

{\bf Discussion.}
We have experimentally unveiled the non-Hermitian BBC of a 1D non-Hermitian system without chiral symmetry implemented in discrete-time non-chiral non-unitary quantum walks of single photons. Through constructing the domain wall structure, we experimentally demonstrated that the non-Hermitian BBC can be faithfully captured by our introduced non-chiral non-Bloch invariants pair. Our experimental observation of the non-Hermitian BBC and the theoretical elucidation of its mechanism are of fundamental importance for understanding topological phenomena in non-Hermitian systems with chiral symmetry breaking. In prospect, our approach should be valuable for advancing the understanding of topological phenomena in open systems.

{\bf  Methods}

{\bf Experimental implementation.}
Our experimental setup is sketched in Fig.~\ref{fig:fig2}. Quantum walks are implemented via the spatial-mode multiplexing scheme. The single-photon source is generated through pumping a $\beta$-barium-borate (BBO) nonlinear crystal via a CW diode laser. Photon pairs at 810nm are produced through the type-I spontaneous parametric down-conversion (SPDC) process. Using one photon as the trigger, the signal photon is heralded in the quantum-walk interferometric network. The signal photon can be prepared in an arbitrary linear polarization state via a polarizing beam splitter (PBS) and wave plates, which then proceeds through the interferometric network. The polarization-dependent translation operator is constructed using the spatial mode, where photons in $\vert 1 \rangle$ are directly transmitted and those in $\vert 0 \rangle$ undergo a lateral displacement into the neighboring spatial mode when passing through beam displacers (BDs), respectively. The coin operator is implemented through half-wave plates (HWPs), which can provide a careful control over parameters ($\theta^{L,R}_{1}, \theta^{L,R}_{2}$). The polarization selective-loss operator $M_{E}=\mathbb{I}_{x} \otimes (\vert 0 \rangle \langle 0 \vert + \sqrt \eta \vert 1 \rangle \langle 1 \vert)$ is implemented by a partially polarizing beam splitter (PPBS), where transmissivities of PPBS are $(H, V)=(1, \eta)$ for horizontally and vertically polarized photons, respectively. Since $M=e^{\gamma}M_{E}$ with $\gamma=-\frac{1}{4} \ln \eta$, it is straightforward to map the experimentally implemented dynamics to those under $U_{0}$ by multiplying a time dependent factor $e^{\gamma t}$. Photons are detected by avalanche photodiodes (APDs) in coincidence with the trigger. The probability distribution of the photon is thus given by the photon counts.

{\bf The split-step quantum walk.}
In our studied non-chiral non-unitary quantum walk, we implement the so-called split-step quantum walk scheme~\cite{Kitagawa2012, Kitagawa2010}. The corresponding Floquet operator can be expressed as $U_{0}=T_{\downarrow}R_{y}(\theta_{2})MT_{\uparrow}R_{y}(\theta_{1})$. Here, the pseudospin-dependent shift operation $T_{\uparrow}(T_{\downarrow})$ can be defined as $T_{\uparrow}=\sum_{j}{\vert j \rangle \langle j-1 \vert \otimes \vert \uparrow \rangle \langle \uparrow \vert + \mathbb{I}_{x} \otimes \vert \downarrow \rangle \langle \downarrow \vert}$ and $T_{\downarrow}=\sum_{j}{\vert j \rangle \langle j+1 \vert \otimes \vert \downarrow \rangle \langle \downarrow \vert + \mathbb{I}_{x} \otimes \vert \uparrow \rangle \langle \uparrow \vert}$, with $j$ being the site index and $\mathbb{I}_{x}=\sum_{j}{\vert j \rangle \langle j \vert}$ being the identity matrix in lattice modes. $\uparrow$ and $\downarrow$ represent the coin state $\vert 0 \rangle$ and $\vert 1 \rangle$, respectively. $R_{y}(\theta)=\mathbb{I}_{x} \otimes e^{-i \theta \sigma_{y}/2}$ stands for the rotation of the coin state around y-axis. $M$ is the polarization selective-loss operator with $\gamma$ being a tunable parameter.

{\bf Acknowledgements.} This work is supported by the National Key R$\&$D Program of China (2021YFA1401700), NSFC (Grants No. 12474267, 12074305), the Fundamental Research Funds for the Central Universities (Grant No. xtr052023002), the Shaanxi Fundamental Science Research Project for Mathematics and Physics (Grant No. 23JSZ003), Shanghai Municipal Science and Technology Major Project (Grant No. 2019SHZDZX01) and the Xiaomi Young Scholar Program. We also thank the HPC platform of Xi'an Jiaotong University, where our numerical calculations were performed.

\onecolumngrid

\renewcommand{\thesection}{S-\arabic{section}}
\setcounter{section}{0}  
\renewcommand{\theequation}{S\arabic{equation}}
\setcounter{equation}{0}  
\renewcommand{\thefigure}{S\arabic{figure}}
\setcounter{figure}{0}  

\indent

\begin{center}\large
\textbf{Supplementary Information}
\end{center}

\section{The bulk of non-chiral non-unitary quantum walks governed by $U_{0}$}
To study the bulk of our 1D non-chiral non-unitary quantum walk captured by
$U_{0}$, utilizing the developed non-Bloch band theory, we rewrite the bulk
eigenstate $\left\vert \psi\right\rangle $ as%
\begin{equation}
\left\vert \psi\right\rangle =\sum\limits_{j,n}\beta_{n}^{j}\left\vert
j\right\rangle \otimes\left\vert \phi_{n}\right\rangle,
\end{equation}
where ${\beta_{n}}$ is the spatial-mode function for the $n$-th mode.
${\phi_{n}}$ is the corresponding coin state and $j$ represents the $j$-th
site of 1D lattice. The bulk of our studied non-chiral non-unitary quantum
walk governed by $U_{0}$ can be investigated through solving the following
eigenstate equation%
\begin{equation}
(A_{m}\beta_{n}+A_{p}\beta_{n}^{-1}+A_{s}-\lambda)\left\vert \phi
_{n}\right\rangle =0,
\end{equation}
where $A_{m}=F_{m}MG_{s}$ , $A_{p}=F_{s}MG_{p}$ and $A_{s}=F_{s}MG_{s}%
+F_{m}MG_{p}$ with $F_{m}=P_{\downarrow}R_{y}(\theta_{2})$ , $F_{s}%
=P_{\uparrow}R_{y}(\theta_{2})$ , $G_{s}=P{_{\downarrow}}R_{y}(\theta_{1})$,
$G_{p}=P_{\uparrow}R_{y}(\theta_{1})$, $P_{\uparrow}=\left\vert \uparrow
\right\rangle \left\langle \uparrow\right\vert $, $P_{{\downarrow}%
}=\left\vert \downarrow\right\rangle \left\langle \downarrow\right\vert $ and
$\lambda=\exp(-iE)$. To make the above equation have non-trivial solutions,
the following relation should be satisfied
\begin{equation}
\det(A_{m}\beta_{n}+A_{p}\beta_{n}^{-1}+A_{s}-\lambda)=0.
\end{equation}
Through solving Eq. (S3), we find that there exist two solutions of ${\beta
_{n}}$ with $n=1,2$ and Eq. (S3) can thus be rewritten as%
\begin{equation}
(A_{m}\beta_{1}+A_{p}\beta_{1}^{-1}+A_{s}-\lambda)\left\vert \phi
_{1}\right\rangle =0,
\end{equation}%
\begin{equation}
(A_{m}\beta_{2}+A_{p}\beta_{2}^{-1}+A_{s}-\lambda)\left\vert \phi
_{2}\right\rangle =0.
\end{equation}
The energy spectrum can thus be obtained as
\begin{equation}
E\equiv E_{\pm}=\pm\arccos(-\frac{\alpha+\alpha^{-1}}{2}\sin\theta_{1}%
/2\sin\theta_{2}/2+\frac{\beta_{n}\alpha+\beta_{n}^{-1}\alpha^{-1}}{2}%
\cos\theta_{1}/2\cos\theta_{2}/2).
\end{equation}
After applying two unitary transformations $\widetilde{S}=\exp(i\theta
_{1}/4\sigma_{y})$ and $S=\exp(-i\pi/4\sigma_{y})$ to
$U_{0}$, Eq. (S3) can be rewritten, which is used to construct the
effective non-Hermitian Hamiltonian in the form of $\mathbf{h\cdot\sigma}$ with%
\begin{equation}%
\begin{split}
h_{x}^{{}}  &  =-\frac{E}{2i\sin E}(\beta_{n}\alpha-\beta_{n}^{-1}\alpha^{-1})\cos
\theta_{2}/2,\\
h_{y}  &  =\frac{E}{2i\sin E}[i(\alpha^{-1}+\alpha)\cos\theta_{1}/2\sin\theta
_{2}/2+i(\beta_{n}\alpha+\beta_{n}^{-1}\alpha^{-1})\sin\theta_{1}/2\cos\theta_{2}/2],\\
h_{z}  &  =\frac{E}{2i\sin E}(\alpha-\alpha^{-1})\sin\theta_{2}/2.
\end{split}
\end{equation}
When considering imposing the OBC, the spatial-mode $\beta_{n}$ should be
chosen along the GBZ. For simplicity, we thus use ${{\beta}}$ to relabel the
spatial-mode and the effective non-Hermitian Hamiltonian ${\bf H}_{\rm eff}$ as shown in Eq. (2) in the main text can be obtained.
\begin{figure}[h]
\includegraphics[width=0.8\textwidth]{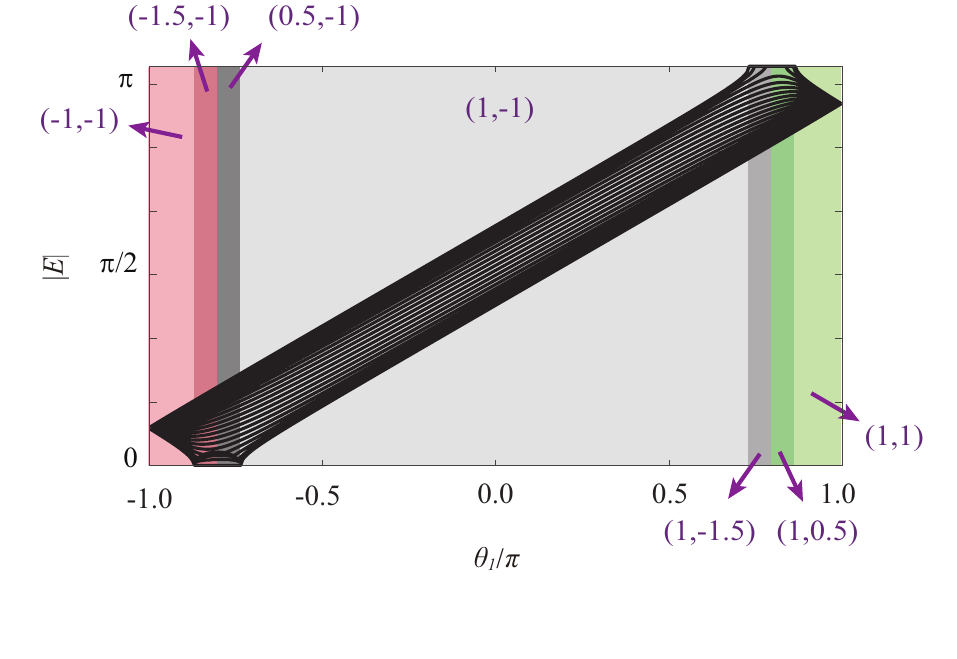}
\caption{{\bf The bulk energy spectrum of the 1D non-chiral non-unitary quantum walk governed by $U_0$ under OBC.
} By fixing the coin parameter $\theta_{2}=0.8\pi$ and varying $\theta_{1}$, it is clearly shown that half integer difference $\bigtriangleup \nu_{0} \ (\bigtriangleup \nu_{\pi})$ between two regions correctly captures the emergence of gap closing
at quasienergy $0 \ (\pi)$ and thus determine the phase boundary between gapped and exceptional regions. Other parameters
are the same as in Fig. 1 in the main text.}
\label{fig:fig3}%
\end{figure}
\section{The non-Bloch theory of domain wall systems}
In our experiments, a domain-wall structure is constructed, where the
interface is built between $j=0$ and $j=-1$, separating the right bulk
$J_{R}(0\leq j\leq N)$ and left bulk $J_{L}(-N\leq j\leq-1)$. Distinct coin
parameters are assigned to the corresponding segments, respectively,%
\begin{equation}
\theta_{1(2)}(j)=\theta_{1(2)}^{L},\text{ }j\in J_{L}%
\end{equation}%
\begin{equation}
\theta_{1(2)}(j)=\theta_{1(2)}^{R},\text{ }j\in J_{R}%
\end{equation}
Under such a domain wall structure, the Floquet operator $U$ can be expressed as%
\begin{equation}
U=\sum\limits_{j}[\left\vert j\right\rangle \left\langle j+1\right\vert
\otimes A_{m}(j)+\left\vert j\right\rangle \left\langle j-1\right\vert \otimes
A_{p}(j)+\left\vert j\right\rangle \left\langle j\right\vert \otimes A_{s}(j)],
\end{equation}
with the site-dependent coin-state operators $A_{m,p,s}(j)$ being defined as%
\begin{equation}%
\begin{split}
A_{m}(j)  &  =F_{m}(j+1)MG_{s}(j+1),\\
A_{p}(j)  &  =F_{s}(j)MG_{p}(j-1),\\
A_{s}(j)  &  =F_{s}(j)MG_{s}(j)+F_{m}(j+1)MG_{p}(j),
\end{split}
\end{equation}
and $F_{m}(j)=P_{\downarrow}R_{y}(\theta_{2}(j))$, $F_{s}(j)=P_{\uparrow}%
R_{y}(\theta_{2}(j))$, $G_{s}(j)=P{_{\downarrow}}R_{y}(\theta_{1}(j))$,
$G_{p}(j)=P_{\uparrow}R_{y}(\theta_{1}(j))$, $P_{\uparrow}=\left\vert
\uparrow\right\rangle \left\langle \uparrow\right\vert $, $P{_{\downarrow}%
}=\left\vert \downarrow\right\rangle \left\langle \downarrow\right\vert $. The
eigenstate of $U$ can be written as $\left\vert \psi\right\rangle =\left\vert
\psi^{R}\right\rangle +\left\vert \psi^{L}\right\rangle $ with%
\begin{equation}
\left\vert \psi^{a}\right\rangle =\sum\limits_{\substack{n, j\in J_{a},a\in
L,R}}\beta_{a,n}^{j}\left\vert j\right\rangle \otimes\left\vert \phi_{n}%
^{a}\right\rangle,
\end{equation}
where $\left\vert \phi_{n}^{a}\right\rangle $ stands for the coin state of the
corresponding bulk and \ $\beta_{a,n}$ is the spatial-mode function. The bulk
of the domain wall structure can be studied through solving the following
eigenstate equation%
\begin{equation}
(A_{m}^{a}\beta_{a,n}+A_{p}^{a}\beta_{a,n}^{-1}+A_{s}^{a}-\lambda)\left\vert
\psi^{a}\right\rangle =0,
\end{equation}
where $A_{m(p,s)}^{a}$ are the corresponding coin operators in the bulk of two
segments, with \ $A_{m(p,s)}^{L}=A_{m(p,s)}^{a}(j)$($-N+1\leq j\leq-2$) and
\ $A_{m(p,s)}^{R}=A_{m(p,s)}^{a}(j)$($1\leq j\leq N-1$). The condition for Eq.
(S13) have non-trivial solutions is
\begin{equation}
\det[A_{m}^{a}\beta_{a,n}+A_{p}^{a}\beta_{a,n}^{-1}+A_{s}^{a}-\lambda]=0.
\end{equation}
Since Eq. (S14) is quadratic in $\beta_{a,n}$, there exist two solutions
$\beta_{a,n}$ with $n=1,2$. Eigenstates of the bulk can thus be rewritten as%
\begin{equation}
\left\vert \psi^{a}\right\rangle =\sum\limits_{n=1,2,j\in J_{a}}\beta
_{a,n}^{j}\left\vert j\right\rangle \otimes\left\vert \phi_{n}^{a}%
\right\rangle.
\end{equation}
We then impose the domain wall boundary conditions at $j=-1$, $j=0$, $j=-N$
and $j=N$. It leads to the following equations \ %
\begin{equation}%
\begin{split}
& \sum\nolimits_{n}(A_{m}^{L\text{ }}\beta_{L,n}^{-N+1}+A_{s}^{L}\beta
_{L,n}^{-N}-\lambda\beta_{L,n}^{-N})\left\vert \phi_{n}^{L}\right\rangle =0,\\
& \sum\nolimits_{n}(A_{p}^{R}\beta_{R,n}^{N-1}+A_{s}^{R}\beta_{R,n}^{N}%
-\lambda\beta_{R,n}^{N})\left\vert \phi_{n}^{R}\right\rangle =0, \\
& \sum\nolimits_{n}[A_{m}^{R\text{ }}\left\vert \phi_{n}%
^{R}\right\rangle +(A_{p}^{L}\beta_{L,n}^{-2}+A_{s}^{L}\beta_{L,n}%
^{-1}-\lambda\beta_{L,n}^{-1})\left\vert \phi_{n}^{L}\right\rangle ]  =0,\\
& \sum\nolimits_{n}[(A_{m}^{R\text{ }}\beta_{R,n}+A_{s}^{R}-\lambda)\left\vert \phi_{n}^{R}\right\rangle +A_{p}^{L}\beta_{L,n}%
^{-1}\left\vert \phi_{n}^{L}\right\rangle ] =0.
\end{split}
\end{equation}
From the above equation, we can derive a set of linear equations $M[\left\vert \phi_{1}^{L}\right\rangle
,\left\vert \phi_{2}^{L}\right\rangle ,\left\vert \phi_{1}^{R}\right\rangle
,\left\vert \phi_{2}^{R}\right\rangle ]^{T}=0$ with $M$ being defined as%
\begin{equation}
M=\left(
\begin{array}
[c]{cccc}%
-A_{m}^{L\text{ }} & -A_{m}^{L\text{ }} &
A_{m}^{R\text{ }} & A_{m}^{R\text{ }}\\
-A_{p}^{L}\beta_{L,1}^{-1} & -A_{p}^{L}\beta_{L,2}^{-1} & A_{p}^{R}\beta
_{R,1}^{-1} & A_{p}^{R}\beta_{R,2}^{-1}\\
A_{p}^{L}\beta_{L,1}^{-N-1} & A_{p}^{L}\beta_{L,2}^{-N-1} & 0_{2\times2} &
0_{2\times2}\\
0_{2\times2} & 0_{2\times2} & A_{m}^{R}\beta_{R,1}^{N+1} & A_{m}^{R}%
\beta_{R,2}^{N+1}%
\end{array}
\right).
\end{equation}
In the thermodynamic limit $N\rightarrow\infty$, the condition of the
existence of non-trivial solutions is that the $8$-by-$8$ coefficient matrix $M$
satisfies $\det(M)=0$. Such a condition can be further simplified as%
\begin{equation}
\zeta(\beta_{a,n})=0,
\end{equation}
with%
\begin{equation}
\zeta(\beta_{a,n}):=\left\{
\begin{array}
[c]{c}%
\left\vert \beta_{L,1}\right\vert -\left\vert \beta_{L,2}\right\vert ,\text{
}|\beta_{R,1}\beta_{L,2}|\geq|\beta_{L,1}\beta_{R,2}|.\\
\left\vert \beta_{R,1}\right\vert -\left\vert \beta_{R,2}\right\vert ,\text{
}|\beta_{L,1}\beta_{R,2}|\geq|\beta_{R,1}\beta_{L,2}|.
\end{array}
\right.
\end{equation}
Through solving Eq. (S18), the bulk energy spectum of the domain wall structure
can be obtained.
\section{The non-chiral non-Bloch topological invariants pair}
To correctly capture the topological properties of our
non-chiral non-unitary quantum walk, the non-chiral non-Bloch
invariants pair is introduced. To define these topological invariants, we rewrite
Floquet operator $U_{0}$ in two different time frames. One of them is denoted by
$U^{\prime}$, which can be expressed as%
\begin{equation}
U^{\prime}=R_{y}(\theta_{1}/2)T_{\downarrow}R_{y}(\theta_{2})MT_{\uparrow
}R_{y}(\theta_{1}/2).
\end{equation}
After applying unitary transformations to $U^{\prime}$, the effective
non-Hermitian Hamiltonian in the form of $\mathbf{h}^{\prime}\mathbf{\cdot
\sigma}$ can be obtained following the same method as treating $U_{0}$, which can be written as
\begin{equation}%
\begin{split}
h_{x}^{\prime}  &  =-\frac{E}{2i\sin E}(\alpha\beta_{n}-\alpha^{-1}\beta_{n}^{-1}%
)\cos\theta_{2}/2,\\
h_{y}^{\prime}  &  =\frac{E}{2i\sin E}[i(\alpha\beta_{n}+\alpha^{-1}\beta_{n}^{-1}%
)\cos\theta_{2}/2\sin\theta_{1}/2+i(\alpha^{-1}+\alpha)\sin\theta_{2}/2\cos\theta
_{1}/2],\\
h_{z}^{\prime}  &  =\frac{E}{2i\sin E}(\alpha-\alpha^{-1})\sin\theta_{2}/2.
\end{split}
\end{equation}
The corresponding eigenstates can be obtained
\begin{equation}
\left\vert \psi_{\pm}^{\prime R}\right\rangle =\frac{1}{\sqrt{2\cos
(2\Omega^{\prime})}}(e^{i(\phi^{\prime}/2\pm\Omega^{\prime})},\pm e^{i(\phi
^{\prime}/2\mp \Omega^{\prime})}e^{i\theta^{\prime}})^{T},%
\end{equation}%
\begin{equation}
\left\langle \psi_{\pm}^{\prime L}\right\vert =\frac{1}{\sqrt{2\cos
(2\Omega^{\prime})}}(e^{-i(\phi^{\prime}/2\mp \Omega^{\prime})},\pm
e^{-i(\phi^{\prime}/2\pm\Omega^{\prime})}e^{-i\theta^{\prime}}),
\end{equation}
with $\cos(2\Omega^{\prime})=\sqrt{\frac{{h'_{x}}^{2}+{h'_{y}}^2+{h'_{z}}^2}{{h'_{x}}^2+{h'_{y}}^2}}$, $\exp(i\theta^{\prime})=\frac{h_{x}^{\prime}+ih_{y}^{\prime}}{\left\vert
h_{x}^{\prime}+ih_{y}^{\prime}\right\vert }$ and $\tan(2\phi^{\prime})=\frac{(\sin
E_{\pm}-\sin E_{\pm}^{\ast})h_{z}^{\prime}}{\sin E_{\pm}+\sin E_{\pm}^{\ast}}%
$. The non-chiral non-Bloch invariant $\nu^{\prime}$ can thus be defined as%
\begin{equation}%
\begin{split}
\nu^{\prime}  &  =-\frac{1}{\pi}\sum\nolimits_{\pm}\oint_{C_{\beta
}^{\prime\text{inside}}(\Delta\rightarrow0)}d\beta\left\langle \psi_{\pm}^{\prime
L}\right\vert i\partial_{\beta}\left\vert \psi_{\pm}^{\prime R}\right\rangle
\\
&  =\frac{1}{\pi}\oint_{C_{\beta}^{\prime\text{inside}}(\Delta\rightarrow
0)}(d\theta^{\prime}+d\phi^{\prime})\equiv \tilde{\nu}^{\prime}+\tilde{\tilde{\nu}}^{\prime},
\end{split}
\end{equation}
where as shown in Fig. S2, $C_{\beta}^{\prime\text{inside}}$ stands for the
corresponding GBZ (inner loop) of $U^{\prime}$ under GBC [30].
\begin{figure}[ptb]
\includegraphics[width=0.88\textwidth]{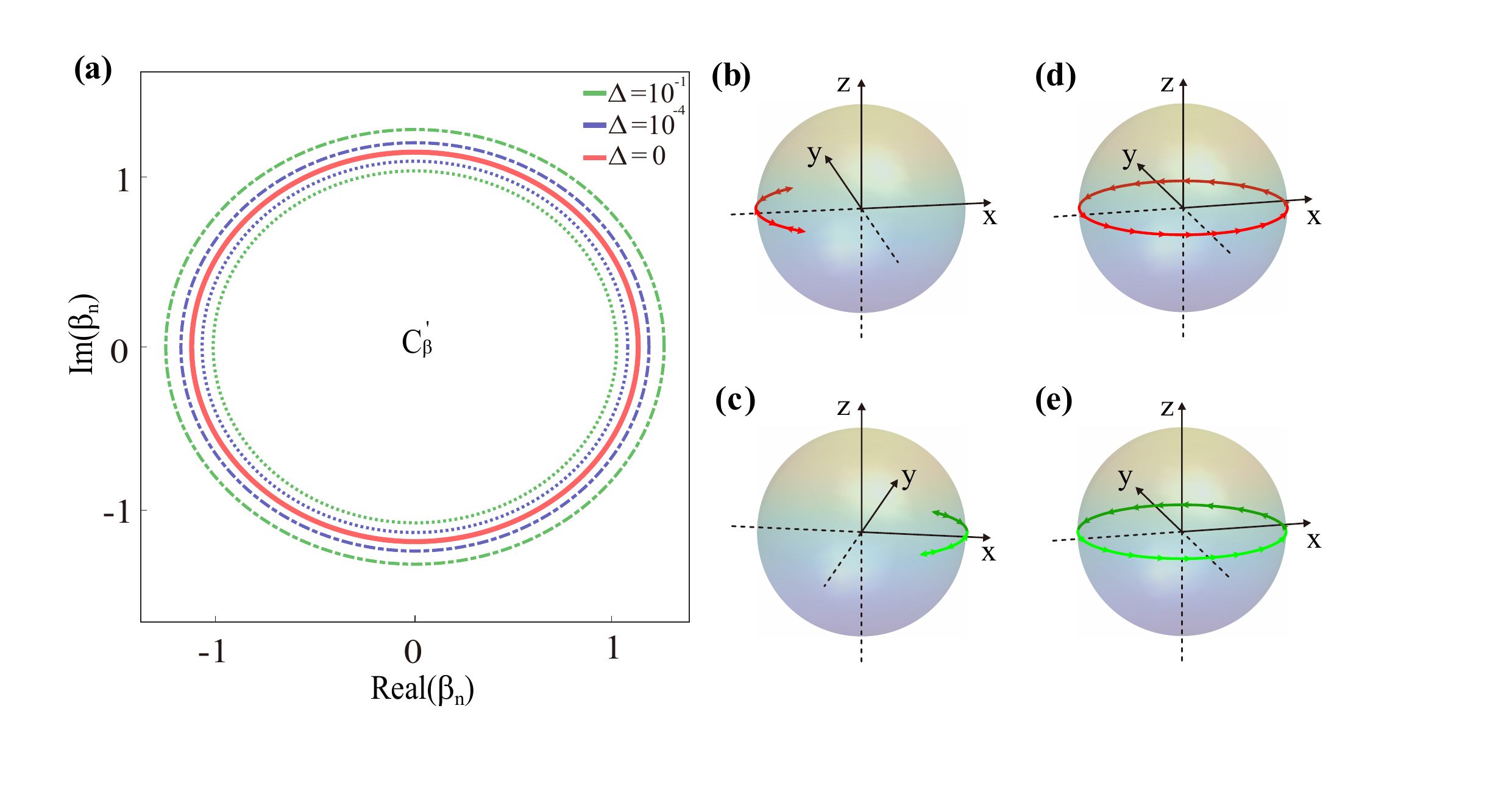}\caption{{\bf The non-chiral non-Bloch topological invariants pair.}
(a) The generalized Brillouin zone (GBZ) for the non-chiral non-unitary quantum walk captured by $U'$ under different GBCs. For
each nonvanished $\Delta$, the GBZ under different GBCs consists of two circles. One is outside the GBZ when  $\Delta=0$ (red circle) and the
other is inside that. Coin parameters are $\theta_{1}=0.95\pi$ and $\theta_{2}=0.8\pi$. (b-e) Unit sphere vector $R_{\pm}$ (red and green curves) capture the corresponding eigenvectors $\left\vert \psi_{\pm}^{^{\prime}R}(\Phi_{\pm}^{\prime},\varphi^{\prime
})\right\rangle$ of $U^{\prime}$. In (b) and (c), coin parameters are chosen as $\theta_{1}=0.3\pi$ and $\theta_{2}=0.5\pi$, which is in the region with ($\nu_{0}=1,\nu_{\pi}=-1$). In (d) and (e), coin parameters are chosen as $\theta_{1}=0.7\pi$ and $\theta
_{2}=0.5\pi$, which is in the region with ($\nu_{0}=1,\nu_{\pi}=1$). Other parameters
are the same as in Fig. 1 in the main text.}
\label{fig:fig42}%
\end{figure}
Following the same procedure above, the Floquet operator can also be given in
another alternative time frame
\begin{equation}
U^{\prime\prime}=R_{y}(\theta_{2}/2)MT_{\uparrow}R_{y}(\theta_{1}%
)T_{\downarrow}R_{y}(\theta_{2}/2).
\end{equation}
The effective non-Hermitian Hamiltonian constructed from Eq. (S25) can be expressed in the form $\mathbf{h}^{\prime\prime}\mathbf{\cdot
\sigma}$\textbf{ }with%
\begin{align*}
h_{x}^{\prime\prime}  &  =-\frac{E}{2i\sin E}(\alpha\beta_n-\alpha^{-1}\beta_n^{-1}%
)\cos\theta_{1}/2,\\
h_{y}^{\prime\prime}  &  =\frac{E}{2i\sin E}[i(\alpha\beta_n+\alpha^{-1}\beta_n^{-1}%
)\cos\theta_{1}/2\sin\theta_{2}/2+i(\alpha^{-1}+\alpha)\sin\theta_{1}/2\cos\theta
_{2}/2],\\
h_{z}^{\prime\prime}  &  =-\frac{E}{2i\sin E}(\alpha-\alpha^{-1})\sin\theta_{1}/2.
\end{align*}
The non-chiral non-Bloch invariant  $\nu^{\prime\prime}$ can thus be obtained through the following definition%
\begin{equation}%
\begin{split}
\nu^{\prime\prime}  &  =-\frac{1}{\pi}\sum\nolimits_{\pm}\oint_{C_{\beta
}^{\prime\prime\text{inside}}(\Delta\rightarrow0)}d\beta\left\langle \psi
_{\pm}^{\prime\prime L}\right\vert i\partial_{\beta}\left\vert \psi_{\pm
}^{\prime\prime R}\right\rangle \\
&  =\frac{1}{\pi}\oint_{C_{\beta}^{\prime\prime\text{inside}}(\Delta
\rightarrow0)}(d\theta^{\prime\prime}+d\phi^{\prime\prime})\equiv\tilde{\nu}^{\prime\prime}+\tilde{\tilde{\nu}}^{\prime\prime},
\end{split}
\end{equation}
where $\exp(i\theta^{\prime\prime})=\frac{h_{x}^{\prime\prime}+ih_{y}%
^{\prime\prime}}{\left\vert h_{x}^{\prime\prime}+ih_{y}^{\prime\prime
}\right\vert }$ , $\tan(2\phi^{\prime\prime})=\frac{(\sin E_{\pm}-\sin E_{\pm
}^{\ast})h_{z}^{\prime\prime}}{\sin E_{\pm}+\sin E_{\pm}^{\ast}}$ and
$C_{\beta}^{\prime\prime\text{inside}}$ refers to the corresponding GBZ
(inner loop) of $U^{\prime\prime}$ under GBC. Therefore,
the non-chiral non-Bloch topological invariant pair can be constructed as
\begin{equation}%
\begin{split}
\nu_{0}  &  =\frac{\nu^{\prime}+\nu^{\prime\prime}}{2},\\
\nu_{\pi}  &  =\frac{\nu^{\prime}-\nu^{\prime\prime}}{2}.%
\end{split}
\end{equation}
As shown in Fig. S1, we find that half integer difference $\Delta \nu_{0}$
($\Delta \nu_{\pi}$) between two regions correctly captures the emergence of  gap
closing at quasienergy $0$ ($\pi$). Our defined non-chiral non-Bloch
topological invariant pair can thus faithfully describe the corresponding gap
closing and determine the phase boundary between full gapped and exceptional regions.

\section{The topological nature of defined non-chiral non-Bloch invariants}

As shown in Eq. (S24) and Eq. (S26), our defined topological invariant can be
expressed into two parts. In experiments, the existence of topological $0$
($\pi$)- edge modes at the interface of domain-wall system can be correctly predicted by
the difference of the part of topological invariant associated with $\theta'$($\theta''$).
In this section, we will show that it can be understood from the geometrical
meaning of the topological invariant associated with $\theta'$($\theta''$). To be more specific,
we monitor the trajectories of eigenvectors by projecting them
onto a 2D unit spherical surface. Taking eigenvectors of $U^{\prime}$ as an
example, the right eigenvector can be parametrized as%

\begin{equation}
\left\vert \psi_{\pm}^{^{\prime}R}\right\rangle =(\cos\frac{\Phi_{\pm}^{\prime}}{2},e^{i\varphi^{\prime}}%
\sin\frac{\Phi_{\pm}^{\prime}}{2})^{T},%
\end{equation}
with $\tan\frac{\Phi_{\pm}^{\prime}}{2}=\frac{\left\vert h_{x}^{\prime}%
+ih_{y}^{\prime}\right\vert }{\left\vert E_{\pm}^{\prime}+h_{z}^{\prime
}\right\vert }$ and $\exp(i\varphi^{\prime})=\frac{(h_{x}^{\prime}%
+ih_{y}^{\prime})\left\vert (E_{\pm}^{\prime\ast}+h_{z}^{\prime\ast
})\right\vert }{\left\vert (h_{x}^{\prime}+ih_{y}^{\prime})\right\vert
(E_{\pm}^{\prime\ast}+h_{z}^{\prime\ast})}$. For each eigenvector $\left\vert
\psi_{\pm}^{^{\prime}R}(\Phi_{\pm}^{\prime},\varphi^{\prime})\right\rangle $,
sphere vector can be defined as $R_{\pm}^{\prime}=(\cos\varphi^{\prime
}\sin\Phi_{\pm}^{\prime},\sin\varphi^{\prime}\sin\Phi_{\pm}^{\prime},\cos
\Phi_{\pm}^{\prime})$, where $\varphi^{\prime}$and $\Phi_{\pm}^{\prime}$
correspond to the azimuthal and polar angles, respectively. In Fig. S2, we
show the trajectories of two eigenvectors on the Bloch sphere across the GBZ, which
form closed curves. When considering distinct regions with
($\nu_{0}=1,\nu_{\pi}=1$) in Fig. S2 (d) and (e) and ($\nu_{0}=1,\nu_{\pi}=-1$) in
Fig. S2 (b) and (c), only $\tilde{\nu}^{\prime}$, the part of topological invariant
associated with $\theta^{\prime}$ in Eq. (S24), is nonzero. The non-chiral non-Bloch invariant
$\nu^{\prime}$ thus satisfy the relation $\nu^{\prime}=\tilde{\nu}^{\prime}$. In the regions with ($\nu_{0}=1,\nu_{\pi}=1$) and ($\nu_{0}=1,\nu_{\pi}=-1$), $\nu^{\prime}=\tilde{\nu}^{\prime}$ are $2$ and $0$, respectively. As shown in
Fig. S2, it is amazing to find that $\tilde{\nu}^{\prime}$ is perfectly consistent with the total  winding number, which accounts for times of the trajectories of two eigenvectors passing around $z$-axis. Therefore, the topological trivial and non-trivial region can be distinguished by the part of topological invariant associated with $\theta^{\prime}$.

\begin{figure}[ptb]
\includegraphics[width=0.6\textwidth]{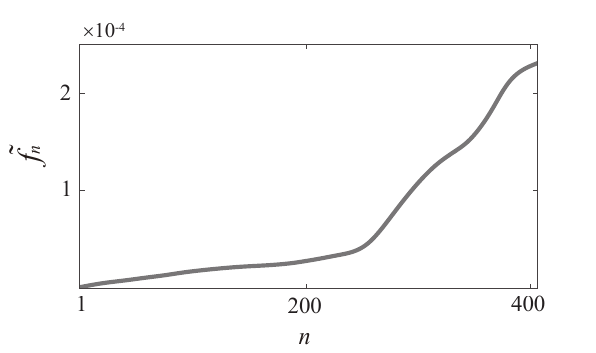}\caption{{\bf Edge states detection scheme under the domain structure.} The ratio $\widetilde{f}_{n}$ compares the contribution between the n-th eigenstate excluding the $\pi$-edge mode and the $\pi$-edge mode. It is clearly shown that the contribution of topological edge $\pi$-edge mode will dominate over others. Here, coin parameters for the domain configuration are ($\theta^{L}_{1}=0.95\pi$, $\theta^{L}_{2}=0.8\pi$) and ($\theta^{R}_{1}=-0.3\pi$, $\theta^{R}_{2}=0.5\pi$). We choose the evolution time $t=100$ and the lattice length $L=203$. Other parameters are the same in Fig. 1 in the main text.}%
\label{fig:fig5}%
\end{figure}

\section{Edge states detection scheme}
In experiments, we utilize the method of weighted summation [26] of the
time-integrated wave function to detect the topological edge state. The
time-dependent wave function of the photon  can be expressed as%
\begin{equation}%
\begin{split}
\left\vert \phi(t)\right\rangle  &  =U^{t}\left\vert \phi(0)\right\rangle
\\
&  =\sum\nolimits_{n}e^{-iE_{n}t}\Phi_{n}\left\vert \psi_{n}\right\rangle,
\end{split}
\end{equation}
where $\Phi_{n}=\left\langle \chi_{n}|\phi(0)\right\rangle $ with $\phi(0)$
being the initial state. $\left\vert \psi_{n}\right\rangle $ \ and
$\left\langle \chi_{n}\right\vert $ are the right and left eigenvector of $U$, respectively. $E_{n}$ labels the corresponding eigenenergy. In the following, we will show that the weighted summation of the time-integrated wave functions
can extract the topolgical edge mode. Let us take the $\pi$-edge mode as an
example. The weighted summation of the time-integrated wave functions can be
written as%
\begin{equation}%
\begin{split}
\left\vert \Phi(t)\right\rangle  &  =\sum_{t^{\prime}=0}^{t}%
\frac{e^{i\pi t^{\prime}}}{t+1}\left\vert \phi(t^{\prime})\right\rangle \\
&  =\sum_{n}f(E_{n})\Phi_{n}\left\vert \psi_{n}\right\rangle,
\end{split}
\end{equation}
with $f(E_{n})=\sum_{t^{\prime}=0}^{t}\frac{e^{i\pi t^{\prime}}}%
{t+1}e^{-iE_{n}t'}$. Since the $\pi$-edge mode has the largest imaginary part of the
eigenenergy, in the long-time dynamics the contribution of $f(E_{n})$
from the topological $\pi$-edge mode will dominate over others. As shown in Fig. S3,
we plot the ratio $\widetilde{f}_{n}=\bar{f}(E_{n})/$ $f
(E_{\pi-edge})$, where $\bar{f}(E_{n})$ refers to the contribution of the n-th eigenstate excluding the $\pi$-edge mode. It is clearly shown that the contribution of topological edge $\pi$-edge mode will dominate over others. Therefore, such a topological edge mode
can be selected out through the method of the weighted summation.

\section{Experimental detection}
The time-evolved wave function of the photon $\vert \phi(t)\rangle$ can be constructed through our measured t-th step time-evolved state $\vert \varphi (t) \rangle$ in experiments via the relation $\vert \phi(t)\rangle=e^{\gamma t} \vert \varphi (t) \rangle$, where $\vert \varphi (t) \rangle$ can be expressed as
\begin{equation}
\vert \varphi (t) \rangle = \sum_{j}{ p_{0}(t,j)\vert j \rangle \otimes \vert 0 \rangle + p_{1}(t,j)\vert j \rangle \otimes \vert 1 \rangle }.
\end{equation}
Here, the coefficients $p_{0}(t,j)$ and $p_{1}(t,j)$ are real numbers, since both $U$ and initial states considered here are real. As illustrated in Fig. 2 in the main text, our experiments involve four kinds of measurements to reconstruct the time-evolved state $\vert \varphi (t) \rangle$. First, the absolute value of $p_{0(1)}(t,j)$ is measured in the detection unit $M_1$. In $M_1$, a PBS is used to separate the horizontally and vertically polarized photons. The photon coincidences measured by APDs are denoted as $N_{H}(t, j)$ and $N_{V}(t, j)$, when the angle of HWP (H$_1$) is set at 0 and $\pi$, respectively. Then, we can determine $\vert p_{0(1)}\vert$ through the following relation $\vert p_{0(1)}\vert=\sqrt{P_{0(1)}(t,j)}$, with
\begin{equation}
\begin{aligned}
& P_{0}(t,j)=\frac{N_{H}(t,j)}{\bar{M}},\\
& P_{1}(t,j)=\frac{N_{V}(t,j)}{\bar{M}},\\
\end{aligned}
\label{eq:h}
\end{equation}
where $\bar{M}$ represents the total coincidence counts input to the initial state preparation.

Second, the relative sign between $p_{0}(t,j)$ and $p_{1}(t,j)$ can be determined in the detection unit $M_{2}$. After passing through a PBS, photon coincidences are denoted as $N_{+}(t,j)$ and $N_{-}(t,j)$ for setting the angle of $H_{1}$ at $+\pi/2$ and $-\pi/2$, respectively. Then, the relative sign between $p_{0}(t,j)$ and $p_{1}(t,j)$ can be decided by the relation
\begin{equation}
2p_{0}(t,j)p_{1}(t,j)=\frac{N_{+}(t,j)-N_{-}(t,j)}{\bar{M}}.
\end{equation}

Third, to reconstruct the wavefunction, we also need to determine the relative sign of coefficients between neighbor sites, i.e., $p_{0(1)}(t, j)$ and $p_{0(1)}(t, j \pm 1)$. As shown in the detection unit $M_{3}$, an addition BD is utilized to combine photons in neighbor sites. After passing through H$_{2}$ at $\pi$ and BD, the horizontally polarized photons in the spatial mode $j-1$ and the vertically polarized photons in the spatial mode $j$ are combined. The photon coincidences in the projective measurement via H$_{3}$ and the following PBS are denoted as $\tilde{N}_{+}$ and $\tilde{N}_{-}$ for the angle of H$_{3}$ set at $+\pi/2$ and $-\pi/2$, respectively. Then, we have
\begin{equation}
2p_{0}(t,j)p_{1}(t,j-1)= \frac{\tilde{N}_{+}- \tilde{N}_{-}}{\bar{M}}.
\end{equation}

Finally, utilizing the detection unit $M_{4}$, the global sign of $p_{0(1)}(t, j)$ with respect to the reference photons reflected by a PBS in the  initial preparation can be determined. For an arbitrary position $j_{w}$ at
each time step, the relative sign between the amplitudes of the reference photons and the walker photons at $j_{w}$ after $t$ steps can be measured in $M_{4}$ by setting the angles of $H_{4}$ and $H_{5}$ at $\pi$. After passing through the PBS, the photon coincidences are labeled as $N_{+}(t,j_{w})$ and $N_{-}(t,j_{w})$ for the angle of $H_{6}$ at $+\pi/2$ and $-\pi/2$, respectively. Then, we obtain
\begin{equation}
2ap_{0}(t,j_{w})=\frac{N_{+}(t,j_{w})-N_{-}(t,j_{w})}{\bar{M}}.
\end{equation}
Therefore, through the above measurement produces, $\vert \varphi (t) \rangle$ can be constructed.
\end{document}